\documentclass[a4paper,fleqn,usenatbib]{mnras}

\usepackage{mathptmx}

\usepackage[T1]{fontenc}
\usepackage{ae,aecompl}

\usepackage{graphicx}	
\usepackage{amsmath}	
\usepackage{amssymb}	

\usepackage{threeparttable}

\usepackage{url}
\usepackage{verbatim}
\usepackage{hyperref}

\def\IRACcolour{$[3.6]$-$[4.5]$~}
\def\chone{$[3.6]$}
\def\chtwo{$[4.5]$}
\def\hboiii{({\sc H}$\beta$ + {\sc [OIII]})}
\def\he{He{\sc II}~}

\def\oiii{{\sc [OIII]}}

\title[Observational constraints for CR7]{No evidence for Population III stars or a Direct Collapse Black Hole in the $z = 6.6$ Lyman-$\alpha$ emitter `CR7'}  
\author[R. A. A. Bowler et al.]{R. A. A. Bowler$^{1}$\thanks{E-mail:
rebecca.bowler@physics.ox.ac.uk}, R. J. McLure$^{2}$, J. S. Dunlop$^{2}$, D. J. McLeod$^{2}$, \and E. R. Stanway$^{3}$, J. J. Eldridge$^{4}$, M. J. Jarvis$^{1,5}$\\
$^{1}$Astrophysics, The Denys Wilkinson Building, University of Oxford, Keble Road, Oxford, OX1 3RH, UK\\
$^{2}$Institute for Astronomy, University of Edinburgh, Royal Observatory, Edinburgh, EH9 3HJ, UK\\  
$^{3}$Department of Physics, University of Warwick, Gibbet Hill Road, Coventry, CV4 7AL, UK\\
$^{4}$Department of Physics, University of Auckland, Private Bag 92019, Auckland, New Zealand\\
$^{5}$Department of Physics, University of the Western Cape, Bellville 7535, South Africa}

\pubyear{2016}

\begin{document}
\label{firstpage}
\pagerange{\pageref{firstpage}--\pageref{lastpage}}
\maketitle

\begin{abstract}
The $z = 6.6$ Lyman-$\alpha$ emitter `CR7' has been claimed to have a Population III-like stellar population, or alternatively, be a candidate Direct Collapse Black Hole (DCBH).
In this paper we investigate the evidence for these exotic scenarios using recently available, deeper, optical, near-infrared and mid-infrared imaging.
We find strong~\emph{Spitzer}/IRAC detections for the main component of CR7 at $3.6\mu {\rm m}$ and $4.5\mu {\rm m}$, and show that it has a blue colour (\IRACcolour $= -1.2\pm 0.3$).
This colour cannot be reproduced by current Pop. III or pristine DCBH models.
Instead, the results suggest that the \chone~band is contaminated by the {\sc [OIII]}$\lambda\,4959,5007$ emission line with an implied rest-frame equivalent width of $EW_{0}$\hboiii$ \gtrsim 2000$\AA.
Furthermore, we find that new near-infrared data from the UltraVISTA survey supports a weaker He{\sc II}$\,\lambda 1640$ emission line than previously measured, with $EW_{0} = 40 \pm 30$\AA.
For the fainter components of CR7 visible in~\emph{Hubble Space Telescope} imaging, we find no evidence that they are particularly red as previously claimed, and show that the derived masses and ages are considerably uncertain.
In light of the likely detection of strong {\sc [OIII]} emission in CR7 we discuss other more standard interpretations of the system that are consistent with the data.
We find that a low-mass, narrow-line AGN can reproduce the observed features of CR7, including the lack of radio and X-ray detections.
Alternatively, a young, low-metallicity ($\sim 1/200\,{\rm Z}_{\sun}$) star-burst, modelled including binary stellar pathways, can reproduce the inferred strength of the \he line and simultaneously the strength of the observed {\sc [OIII]} emission, but only if the gas shows super-solar $\alpha$-element abundances (O/Fe $\simeq 5\,$(O/Fe)$_{\sun}$).

\end{abstract}


\section{Introduction}

The Lyman-$\alpha$ emission line at $\lambda_{0} = 1216\,$\AA~provides a unique probe of the progress and topology of reionization at $z~>~6$~\citep[e.g.][]{Dijkstra2014}.
Using narrow-band surveys it is possible to select large samples of Lyman-$\alpha$ emitting galaxies (LAEs) up to $z \simeq 7$~\citep[e.g.][]{Ouchi2008, Ouchi2010, Matthee2015} and potentially to higher redshifts~\citep[e.g.][]{Tilvi2010, Krug2012}.
Several of these narrow-band selected galaxies at $z = 6.6$ have generated considerable interest due to their particularly strong (${\rm log}_{10}[L_{Ly\alpha}/{\rm ergs}/{\rm s}] > 43$) and extended ($> 10\,{\rm kpc}$) Lyman-$\alpha$ emission.
The low-metallicity, triple-merger system `Himiko' has been extensively studied~\citep{Ouchi2009, Ouchi2013, Zabl2015}, and recently~\citet{Sobral2015} (hereafter S15) reported an even brighter LAE, `CR7', which was found within the degree-scale Cosmic Origins Survey (COSMOS) field.
CR7 was initially discovered during a search for Lyman-break galaxies in~\citet{Bowler2012}, and was independently selected by S15 in Subaru/Suprime-Cam narrow-band imaging (using the $NB921$ filter centred at $9210$\AA).
Follow-up spectroscopy confirmed the presence of a strong Lyman-$\alpha$ emission line with a rest-frame equivalent width in excess of ${EW}_{0} > 200$\AA. 
Near-infrared spectroscopy of CR7 also revealed a $\sim 6\sigma$ emission-line attributed to He{\sc II} $\lambda 1640$.
The \he line was observed to be sufficiently strong to boost the available $J$-band photometry by $0.4$ magnitudes (S15), with an inferred equivalent width of ${EW}_{0} = 80\pm 20$\AA.
The strong and narrow He{\sc II} line, coupled with the non-detection of metal lines in the near-infrared spectrum, has led to the interpretation that this galaxy has a Population III-like stellar population (S15,~\citealp{Pallottini2015},~\citealp{Visbal2016},~\citealp{Xu2016}~\citealp{Yajima2016}) or alternatively harbours an accreting Direct Collapse Black Hole (DCBH;~\citealp{Dijkstra2016, Smith2016, Agarwal2016, Hartwig2016, Smidt2016}).

With high resolution imaging from archival~\emph{HST} data, CR7 appears as three distinct clumps, with the Lyman-$\alpha$ and \he emission peaking at the location of the brightest component (A).
The two fainter objects (B and C) are separated from the A component by $\gtrsim 5\,{\rm kpc}$ (assuming that they are also at $z = 6.6$) and appear redder, leading to the interpretation that they are older (S15).
In recent theoretical models of CR7, components B and C potentially provide the required ionizing photons to suppress star-formation in component A at earlier times, leading to the formation of a pristine Pop. III star-burst (S15) or DCBH~\citep{Agarwal2016}.
In both scenarios, the presence of older companions is required to reproduce the observed rest-frame optical detections (in the~\emph{Spitzer} \chone~and \chtwo~bands), which are not predicted by the extremely blue Pop. III or DCBH spectral energy distributions (SED).
In simulations of potential sites of Pop. III star-formation, it has appeared challenging to recreate the properties of CR7 due primarily to the large mass ($\ge 10^{7}\,{\rm M}_{\sun}$; S15;~\citealp{Visbal2016}) of Pop. III stars required to reproduce the Lyman-$\alpha$ and \he luminosities~\citep{Hartwig2016, Pallottini2015, Xu2016}.
Another possible issue with the Pop. III scenario is the short ($\lesssim 5\,{\rm Myr}$) visibility timescale of such a burst if it occurred~\citep{Hartwig2016}.
While the DCBH interpretation alleviates these concerns somewhat, with visibility timescales of tens of ${\rm Myr}$~\citep{Pallottini2015}, the line luminosities require a system with a mass greater than the maximum mass thought to be created by direct collapse, requiring the pristine DCBH to have accreted substantially~\citep{Dijkstra2016, Smidt2016}.
Given the intense interest in CR7 and the exciting implications of the potential discovery of a Pop. III star-burst or a DCBH, it is important to scrutinise the observational evidence.
While future observations with~\emph{HST} and the Atacama Large Millimeter/Sub-millimeter Array (ALMA) will provide further insights into the properties of CR7, it is possible to discern salient details about the system from currently available broad-band photometric data.

In this work we present an analysis of the most recently available imaging data for CR7, which extends up to $1$ mag deeper than that presented in S15.
Crucially this includes deeper near-infrared data from the third data release of the UltraVISTA survey and deeper~\emph{Spitzer} data at $3.6\mu {\rm m}$ and $4.5\mu {\rm m}$.
The datasets are described in detail in Section~\ref{sect:data}.
Our analysis of the updated photometric data is presented in Section~\ref{sect:results}.
In Section~\ref{sect:discussion} we compare the observed~\emph{Spitzer}/IRAC colours to those predicted by Pop. III and DCBH models,  where we find that neither model can reproduce the data.
In light of our findings, we discuss alternative interpretations for the nature of CR7 in Section~\ref{sect:alternative}, which includes a comparison of the properties of CR7 to the Binary Population and Spectral Synthesis (BPASS) stellar population models~\citep{Eldridge2008, Eldridge2009}.
We end with our conclusions in Section~\ref{sect:conc}.
All magnitudes are quoted in the AB system~\citep{Oke1974, Oke1983}.
At $z = 6.6$, a measured separation of $1\,$arcsec corresponds to a proper distance of $ 5.4\,{\rm kpc}$ assuming a cosmology with $\Omega_{m} = 0.3$, $\Omega_{\Lambda} = 0.7$ and $H_0 = 70\,{\rm km}\,{\rm s}^{-1}\,{\rm Mpc}^{-1}$.

\section{Data}\label{sect:data}

We use ground-based optical data in the $u^*griz$ bands from the Canada-France-Hawaii Telescope Legacy Survey, and~\emph{HST}/Advanced Camera for Surveys (ACS) $I_{814}$ imaging taken as part of the COSMOS survey~\citep{Koekemoer2007}.
Deeper imaging in the $z'$-band from Subaru/Suprime-Cam was also included~\citep{Furusawa2016}.
The ground-based near-infrared imaging analysed was from the third data release (DR3)\footnote{Public data release can be found at \url{http://www.eso.org/sci/observing/phase3/data_releases.html}} of the UltraVISTA survey~\citep{McCracken2012}, which provides imaging in the $YJHK_{s}$ bands.
In the COSMOS field, CR7 is located close to the edge of one of the `ultra'-deep strips of the UltraVISTA survey, and hence has shallower depths by $0.2$--$0.4$ mag than for these deeper regions quoted in~\citet{Bowler2016}.
In addition to the ground-based imaging,~\emph{HST}/WFC3 data exists for CR7 in the $YJ_{110}$ and $H_{160}$ bands.
This data was taken as part of an unrelated proposal (PI F{\"o}rster Schreiber, ID 12578).
We independently reduced this data (see section 2.3 of~\citealp{Bowler2016} for details), matching the astrometry to the UltraVISTA imaging with a resulting accuracy of RMS $< 0.1$ arcsec.
Finally, we used new mid-infrared imaging at $3.6\mu {\rm m}$ and $4.5\mu {\rm m}$ (hereafter denoted \chone~and \chtwo) from the~\emph{Spitzer}/Infrared Array Camera (IRAC) taken as part of the~\emph{Spitzer} Large Area Survey with Hyper-Suprime-Cam (SPLASH;~\citealp{Steinhardt2014}).
Further details of the image processing, including astrometric and photometric consistency checks performed on the ground-based datasets (which were also performed for the UltraVISTA DR3 images), can be found in~\citet{Bowler2014}.
Errors on the photometry presented in this work were determined from empty aperture measurements.
Using the segmentation map produced by {\sc SExtractor}, we identified blank regions of the images and measured the flux in circular apertures of the appropriate diameter  ($1.2$--$3$ arcsec) to match that used for the galaxy photometry.
A local depth was then calculated for each point in the field by determining the standard deviation (with the median absolute deviation estimator) of the closest 200 apertures.
This approach accounts for the correlated noise in the images by directly measuring the noise in a given aperture size.

\begin{figure*}
\includegraphics[width = \textwidth]{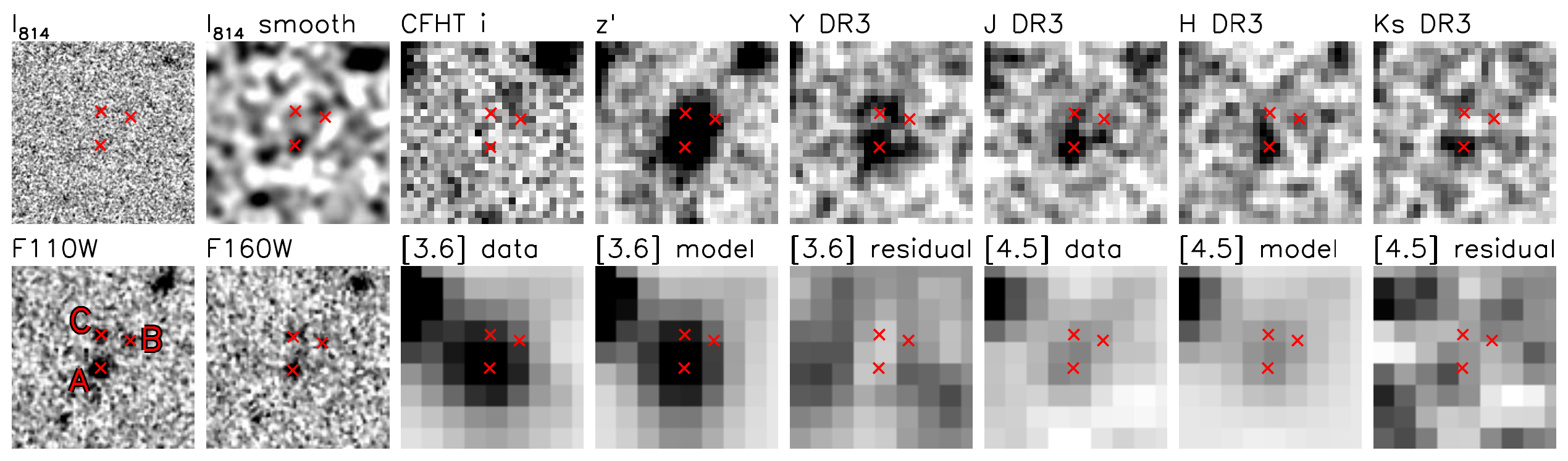}
\caption{Postage-stamp cut-out images for CR7, showing the deconfusion of the~\emph{Spitzer}/IRAC images.
The stamps are $5 \times 5$ arcsec with North to the top and the East to the left.
The red crosses show the positions of the three components visible in the~\emph{HST}/WFC3 data.
The upper row of images from left to right show the~\emph{HST}/ACS $I_{814}$ data, at original resolution and smoothed by a Gaussian kernel ($\sigma = 0.15\,$arcsec), the ground-based $i, z'$ data from the CFHTLS and Subaru/Suprime-Cam data, and the UltraVISTA DR3 data in the near-infrared bands ($Y, J,H, K_{s}$).
The lower row of images shows the~\emph{HST}/WFC3 imaging in the $YJ_{110}$ and $H_{160}$ bands on the left.
The final six stamps show the \chone~data, the \chone~model and the residual followed by the similar \chtwo~results.
For the optical and near-infrared stamps, we have saturated to black any pixel that exceed the $3\sigma$ limit (per pixel), and to white if the pixel value is lower than $-1.5\sigma$.
For the IRAC stamps of the data and model, we saturate pixels in the same manner but in the range [$-1\sigma$, $10\sigma$], and for the residual stamps we adjust the range to [$-2\sigma$, $2\sigma$].
}\label{fig:photo}
\end{figure*}

In the selection and further analysis of CR7 presented in~\citet{Matthee2015} and S15, the optical data utilized was from the Subaru/Suprime-Cam imaging of the COSMOS field~\citep{Taniguchi2007}, the near-infrared data was from the second data release (DR2) of UltraVISTA survey and the~\emph{Spitzer}/IRAC imaging was from S-COSMOS~\citep{Sanders2007}.
In comparison to S15, the data utilized in this work is $0.3$--$0.6$ mag deeper in the near-infrared $YJHK_s$ filters and $\gtrsim 1.0$ mag deeper in the~\emph{Spitzer} bands.

\section{Results}\label{sect:results}

Postage-stamp cutout images of CR7 from a selection of the available optical to mid-infrared imaging is shown in Fig.~\ref{fig:photo}.
As expected for a $z = 6.6$ galaxy, there is no detection in the optical data at the position of any of the components.
In the spectroscopic data, S15 finds a spatially compact, continuum detection corresponding to rest-frame wavelengths of $916$--$1017$\AA~that they attribute to Lyman-Werner photons emitted by CR7.
We find no strong detection in the CFHT $i$-band data, which has a $5\sigma$ limiting depth of $26.7$ ($1.8$ arcsec diameter circular aperture) and is well matched in wavelength to measure such a signal.
There is, however, a weak CFHT $i$-band detection to the North of component B.
The weak detection is measured to be $m_{\rm AB} \simeq 28.5$, which is around a $1.5\sigma$ detection, and could be a very faint foreground galaxy or a spurious source.
The strong $z$-band detection, which is a combination of continuum emission and Lyman-$\alpha$ flux, is peaked at the position of component A (the Pop. III/DCBH candidate) and is clearly extended in the direction of components B and C.

\subsection{Rest-frame UV emission}\label{sect:uv}

\begin{table*}                                                                                     
\caption{The measured photometry for CR7 utilizing deeper near-infrared data from UltraVISTA DR3 and mid-infrared imaging from the SPLASH survey.
The upper part of the table shows the available~\emph{HST} and~\emph{Spitzer} photometry.
For the full object photometry, labelled as component `Full', the photometry was measured in $3$ arcsec diameter circular apertures on the~\emph{HST}/WFC3 data, and the~\emph{Spitzer}/IRAC photometry was obtained using a deconfusion analysis based on the UltraVISTA $K_{s}$-band data.
The lower part of the table shows the ground-based photometry measured in $1.2$, $2.0$ and $3$ arcsec diameter circular apertures.
Note that the aperture photometry has been corrected to total assuming a point-source correction.
The final row of the table reproduces the photometry and $\beta$ value as measured in S15.}
\begin{tabular}{c c c c c c c c c c}
\hline
Component & $I_{814}$ & $I_{814} - YJ_{110}$ & \multicolumn{2}{c}{$YJ_{110}$} & $H_{160}$  & $YJ_{110} - H_{160}$ & $[3.6]$ & $[4.5]$ &  $[3.6] - [4.5]$ \\

 \hline
A  &   $ > 27.2 $  & $ >  2.2 $ &  \multicolumn{2}{c}{$ 25.01_{-0.05}^{+0.05} $}  &  $ 25.25_{-0.12}^{+0.13} $   & $ -0.24_{-0.18}^{+0.17} $  & $23.79_{-0.09}^{+0.10}$ & $24.99_{-0.19}^{+0.23}$ & $-1.20_{-0.32}^{+0.29}$ \\[1ex]
B  &  $ > 28.2 $  &  $ >  1.2 $  & \multicolumn{2}{c}{$ 27.02_{-0.13}^{+0.15} $}  &  $ 26.88_{-0.23}^{+0.30} $  &   $ +0.14_{-0.43}^{+0.38} $  & $25.99_{-0.38}^{+0.59}$ & $25.97_{-0.30}^{+0.41}$ & $+0.03_{-0.79}^{+0.89}$ \\[1ex]
C  &  $ > 28.2 $  & $ >  1.5 $  & \multicolumn{2}{c}{$ 26.67_{-0.09}^{+0.10} $}  &  $ 26.80_{-0.22}^{+0.27} $  &  $ -0.13_{-0.37}^{+0.32} $  & $24.97_{-0.26}^{+0.34}$ & $25.56_{-0.32}^{+0.45}$ & $-0.59_{-0.71}^{+0.65}$ \\[1ex]
Full & & & \multicolumn{2}{c}{$24.65_{-0.07}^{+0.08}$} &$24.70_{-0.15}^{+0.18}$& & $23.49_{-0.09}^{+0.10}$ & $24.57_{-0.12}^{+0.13}$ &$-1.08_{-0.22}^{+0.22}$\\
\hline
Aperture & $i$ & $z'$ & $Y$ & $J$ & $H$ & $K_{s}$ & $\beta_{YHK}$ & Notes & \\
\hline

$1.2$ arcsec & $ > 28.42 $ & $ 25.24_{- 0.10}^{+ 0.11} $ & $ 25.09_{- 0.11}^{+ 0.12} $ & $ 24.91_{- 0.13}^{+ 0.14} $ & $ 25.10_{- 0.17}^{+ 0.20} $ & $ 24.96_{- 0.22}^{+ 0.28} $&$-1.9_{-0.3}^{+0.3}$&
\multicolumn{2}{l}{Component A} \\[1ex]
$2.0$ arcsec & $ > 27.54 $ & $ 25.00_{- 0.10}^{+ 0.11} $ & $ 24.71_{- 0.10}^{+ 0.11} $ & $ 24.63_{- 0.12}^{+ 0.14} $ & $ 25.01_{- 0.23}^{+ 0.28} $ & $ 25.16_{- 0.26}^{+ 0.34} $&$-2.5_{-0.4}^{+0.3}$&
\multicolumn{2}{l}{Full system} \\[1ex]
$3.0$ arcsec & $ > 27.10 $ & $ 24.90_{- 0.10}^{+ 0.11} $ & $ 24.75_{- 0.16}^{+ 0.18} $ & $ 24.68_{- 0.18}^{+ 0.21} $ & $ 24.92_{- 0.31}^{+ 0.43} $ & $ 25.12_{-0.44}^{+0.76} $&$-2.4_{-0.6}^{+0.5}$& \multicolumn{2}{l}{Full system} \\
\hline
$2.0$ arcsec & $ - $ & $ 25.35_{-0.20}^{+0.20} $ & $ 24.92_{-0.13}^{+0.13} $ & $ 24.62_{-0.10}^{+0.10} $ & $ 25.08_{-0.14}^{+0.14} $ & $ 25.15_{-0.15}^{+0.15} $&$-2.3_{-0.08}^{+0.08}$&
\multicolumn{2}{l}{\citet{Sobral2015}} \\
\hline

\end{tabular}

\label{table:cr7}
\end{table*}

In the UltraVISTA DR3 imaging, CR7 is clearly detected in all four filters, and appears spatially extended in the deeper $Y$ and $J$-bands. In this data, the brightest component A clearly dominates and is visible in the shallower $H$ and $K_s$ stamps.
Components B and C are also visible at approximately the $ 2\sigma$ level in the $Y$ and $J$ bands, consistent with the measured magnitudes in the~\emph{HST}/WFC3 imaging.
To determine the photometry and colour of the Pop. III/DCBH candidate, we measured photometry on the ground-based imaging centred on the position of component A using small circular apertures of $1.2$ arcsec in diameter.
We also measured the photometry in larger apertures of $2$ arcsec in diameter to compare directly with the results of S15, and using a diameter of $3$ arcsec to provide a more appropriate measure of the total flux.
The larger apertures were centred on the peak of the $Y+J$ UltraVISTA image, which is offset by $0.35$ arcseconds to the North-West of component A (the stamps shown in Fig.~\ref{fig:photo} are centred on this position).
The results are shown in Table~\ref{table:cr7}.
All photometry was corrected to total magnitudes assuming a point-source correction.\footnote{For the UltraVISTA images we consider, the enclosed flux for a point-source within apertures of diameter $1.2, 2.0$ and $3.0$ arcsec respectively, was $[Y,J,H,K_{s}] = [56, 60, 63, 64]$, $[79, 83, 85, 87]$ and $[92, 94, 95, 96]$ percent.}
S15 found a strong excess in the $J$-band photometry for CR7 using the UltraVISTA DR2 data, when the continuum level was derived from fitting to the $Y$, $H$ and $K_s$ bands.
The observed offset of $\Delta m_{\rm AB} = -0.4 \pm 0.13$ mag was interpreted as the contribution to the measured photometry from the spectroscopically observed \he emission line.
Converting the $J$-band excess into an estimated rest-frame equivalent width, S15 inferred an equivalent width of $EW_{0} = 80 \pm 20$\AA~for the \he line.

\begin{figure}
\includegraphics[width = 0.47\textwidth]{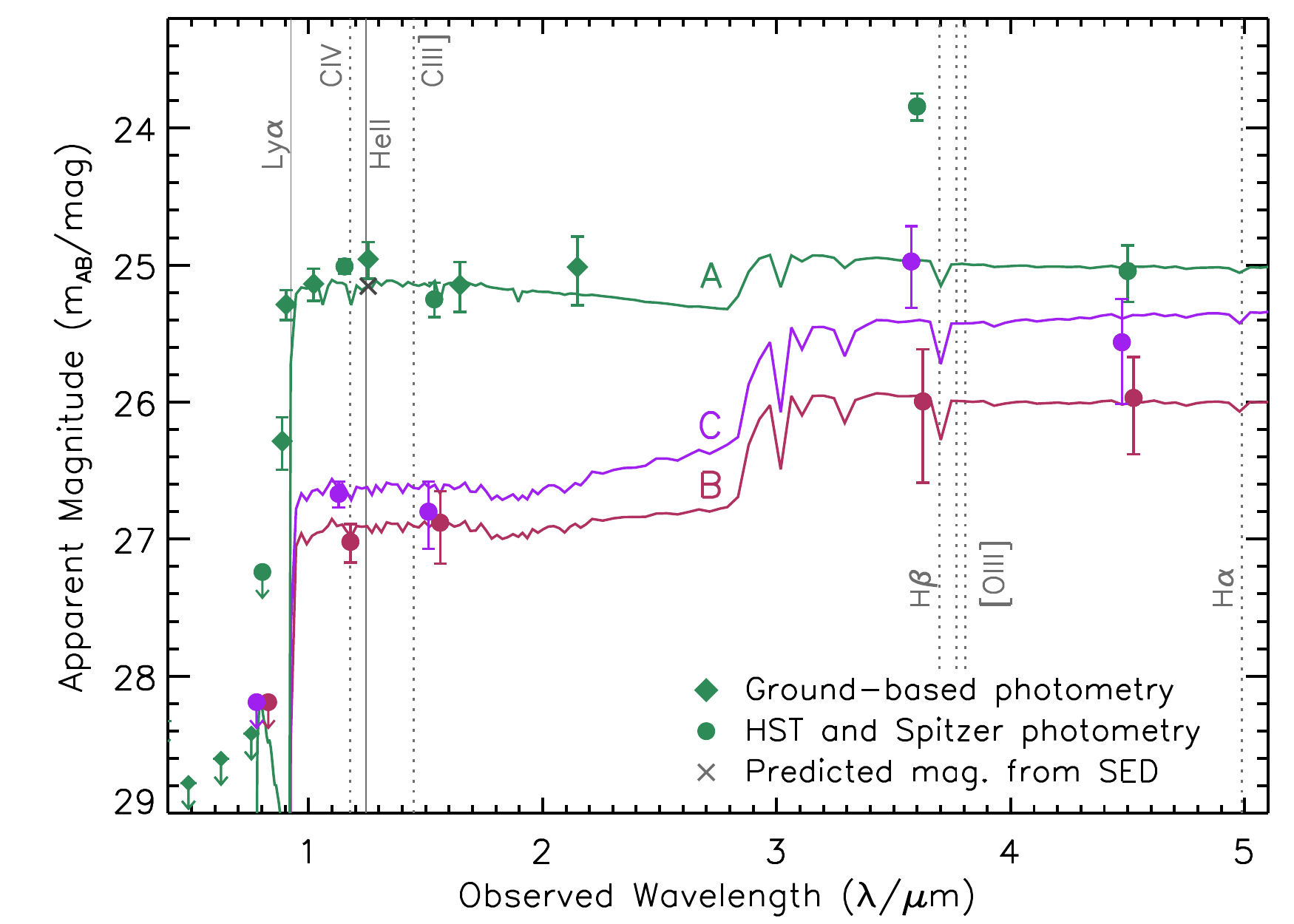}
\caption{The measured photometry for CR7 from the UltraVISTA DR3 imaging and the deconfused SPLASH data.
The~\emph{HST} and~\emph{Spitzer} photometry is shown as the filled circles, for components A (green), C (purple) and B (red) from top to bottom.
The ground-based photometry is shown for the brightest component A, as the filled diamonds.
The best-fitting~\citet{Bruzual2003} model is shown as the solid line in each case, with the predicted model photometry in the $J$-band shown as the black cross (for component A).
The solid vertical grey lines show the observed wavelengths of the confirmed Ly-$\alpha$ and \he emission lines (S15), whereas the dotted vertical lines show for reference the wavelengths of other rest-frame UV and optical lines that have been detected or inferred in other high-redshift LBGs~\citep{Stark2014a, Stark2015, DeBarros2016, Smit2014}.
}\label{fig:sed}
\end{figure}

Using the deeper near-infrared imaging now available, we find a smaller $J$-band excess than that determined in S15, with $\Delta m_{\rm AB} = -0.19_{-0.13}^{+0.14} $ for the $1.2$ arcsec diameter measurement. 
Here the continuum level was determined from the best-fitting~\citet{Bruzual2003} SED to the $Y$, $H$, $K_{s}$ and \chone~bands with the redshift fixed to $z = 6.6$ (as shown in Fig.~\ref{fig:sed}).
If instead the continuum level in the $J$-band is determined from a power-law fit to the $Y$, $H$ and $K_{s}$ bands, we find an excess of $\Delta m_{\rm AB} = -0.17_{-0.13}^{+0.14} $.
An offset of $\simeq 0.2\,{\rm mag}$ was also found using the two larger aperture measurements, and the results are unchanged if PSF homogenised images are used.
To understand the differences between our results and those of S15, we compared photometric catalogues from the older DR2 data used by S15, and the DR3 data.
In the four bands we find no significant zeropoint offset between the two releases, and the aperture magnitudes for CR7 agree within $\lesssim 0.1\,{\rm mag}$ for the $Y$, $H$ and $K_s$ imaging.
In the $J$-band however, we find that the measured magnitude for CR7 in the DR3 data is $0.22\,{\rm mag}$ fainter than in the DR2 imaging.
Such an offset is expected for 10 percent of the objects in this magnitude range due to the photometric errors on the measurement (measured by directly comparing catalogues from the DR2 and DR3 data), and represents a $\sim1.5\sigma$ deviation from the DR2 value.
If we interpret our observed flux excess as due to contamination of this broad band by an emission line (or lines) at $z = 6.6$, the magnitude of the offset implies an $EW_{0} = 40 \pm 30$\AA~(using equation 1 in~\citealp{MarmolQueralto2015} assuming a filter width of 1740\AA;~\citealp{vanderWel2011}).
We note however, that the near-infrared photometry could also be well fitted by a continuum only model, given the errors on the photometry.
The lower $EW_{0}$ we infer is fully consistent with the lower limit calculated from the spectroscopic detection of the likely \he line presented in S15, which implies $EW_{0} > 20$\AA.

Comparing our near-infrared photometry in Table~\ref{table:cr7} to that measured in S15 using an identical aperture ($2.0$ arcsec diameter), we find good agreement except in the $Y$-band, where our measurement is brighter than the value obtained by S15. 
In addition, the comparison shows that the photometric errors presented in S15 were underestimated, leading to the very small errors on the derived $\beta$ value.
It is unclear whether S15 used an aperture correction to account for flux that lies beyond the circular aperture used.
If no aperture correction was applied by S15, this would provide a natural explanation of the differences between our results in the $Y$-band (which shows the largest offset), as the PSF has considerably larger wings in this band than at longer wavelengths and hence requires the largest aperture correction.
The apparent agreement between our results and those of S15 in the $J$-band is then a result of our aperture correction approximately matching the $\sim 0.2\,$ mag offset we find between the DR2 and DR3 photometry (measured from raw, uncorrected aperture photometry).
Comparing the S15 photometry to the PSF homogenised results from the COSMOS15 catalogue~\citep{Laigle2016} supports this interpretation, as the S15 magnitudes are only in good agreement after a wavelength dependent aperture correction is applied.
Our results are in excellent agreement with those measured in the COSMOS15 catalogue, which uses the UltraVISTA DR2 data, except in the $J$-band where we again find an offset of around $0.2\,$mag as compared to the DR3 photometry.
The $J$-band excess measured from the COSMOS15 catalogue implies a \he equivalent width of $EW_0 = 100^{+50}_{-40}\,$\AA, which is again approximately double the value we derive from the deeper UltraVISTA DR3 data. 

We measured the rest-frame UV slope, $\beta$ ($F_{\lambda} \propto \lambda^\beta$) of CR7 by fitting a power law to the $Y$, $H$ and $K_{s}$-band photometry.
The error was calculated from the resulting $\chi^2$ distribution.
Even with the deeper UltraVISTA DR3 data, we find the $\beta$ value to be considerably uncertain.
At the position of component A, using small $1.2$ arcsec diameter circular apertures, we find a rest-frame UV slope that is consistent with that found for normal LBGs at high-redshift ($\beta = -1.9^{+0.3}_{-0.3}$;~\citealp{Bouwens2014beta, Dunlop2012a, Dunlop2013}).
S15 found a bluer value of $\beta = -2.3 \pm 0.08$, with a considerably smaller error despite the shallower UltraVISTA DR2 data used.
If we use the same aperture ($2$ arcsec in diameter), correcting to total magnitudes, we also find a bluer value more consistent with the results of S15 ($\beta = -2.5_{-0.4}^{+0.3}$) however with more realistic errors (e.g for similarly bright $z > 6$ LBGs;~\citealp{Bowler2015}).
Using the photometry and errors provided for CR7 in the COSMOS15 catalogue, we derive $\beta = -2.4^{+0.4}_{-0.6}$ ($2\,$arcsec aperture).
The three $\beta$ measurements we measure using the UltraVISTA DR3 data (shown in Table~\ref{table:cr7}), are all consistent within the errors, however we find bluer values for the larger aperture measurements.
The slightly bluer values are a result of contamination of the larger aperture photometry in the $Y$-band (which has the most significant PSF wings) by components B and C, which are visible in the UltraVISTA DR3 $Y$-band imaging as extended emission.

We also measured the photometry and resulting colours for the three individual components from the~\emph{HST}/WFC3 imaging.
Apertures of diameter of $1$, $0.4$ and $ 0.4$ arcsec were used for components A, B and C respectively, correcting to a total magnitude assuming a point source.
The results are shown in Table~\ref{table:cr7}.
There is considerable error in the derived colours due to the faintness of these components in the $H_{160}$ band, but components B and C show colours consistent with a constant $F_{\nu}$ in the rest-frame UV ($\beta \simeq -2$; Fig.~\ref{fig:sed}).
We therefore find no evidence that components B and C are particularly red as claimed by S15. 
Finally, we note that the rest-frame UV colour of component A cannot be robustly inferred from the~\emph{HST}/WFC3 photometry because of the contamination of the broad $YJ_{110}$ band by both the Lyman-$\alpha$ and likely \he emission lines.

\subsection{Rest-frame optical emission}

The~\emph{Spitzer}/IRAC data analysed in S15 was from the S-COSMOS survey, which is $1$ mag shallower than the SPLASH imaging utilized here.
As shown in Fig~\ref{fig:photo}, CR7 is clearly detected in the \chone~and \chtwo~filters in the SPLASH data and appears to show a blue colour in these bands.
In~\citet{Bowler2014} we analysed the SPLASH data for CR7 (ID 30425) as part of a search for LBGs in the UltraVISTA DR2 data.
We found that CR7 has a strong blue colour of \IRACcolour $= -1.4^{+0.5}_{-0.6}$ mag, consistent with contamination of the~\emph{Spitzer}/IRAC filters by rest-frame optical emission lines (e.g. as found by~\citealp{Smit2014}).
S15 attributed the majority of the IRAC flux detected in the S-COSMOS imaging to the UV faint components B and C, as the Pop. III model fitted to the brighter component A does not significantly contribute to the~\emph{Spitzer} bands.

To investigate further the origin of the rest-frame optical emission in the CR7 system we have improved on the methodology presented in~\citet{Bowler2014} using a deconfusion analysis of the~\emph{Spitzer}/IRAC SPLASH imaging.
We deconfused the SPLASH imaging using {\sc TPHOT}~\citep{Merlin2015}, using the~\emph{HST}/WFC3 $YJ_{110}$ data as the high-resolution input data (our results are unchanged if the $H_{160}$ imaging is used).
The best-fitting model from the deconfusion analysis and the residual are shown for the \chone~and \chtwo~bands in Fig.~\ref{fig:photo}.
We find that the dominant source of the emission observed in these bands is emitted from component A, not the two fainter components, and that this emission shows a strong blue colour of \IRACcolour $ = -1.2 \pm 0.3$ (see Table~\ref{table:cr7}).
Our results are in good agreement with those presented in~\citet{Agarwal2016}, who similarly found that 70 percent of the IRAC emission emanates from component A using a similar methodology.
The \IRACcolour colours measured for components B and C are consistent with zero.
To test the results of our deconfusion, we re-ran the analysis masking component A and hence requiring that components B and C are the only sources of the observed IRAC fluxes.
The result is an extremely poor fit to the observed data, with a residual of order 50 percent of the peak flux at the position of component A.
As we discuss in more detail below, the strong rest-frame optical detection observed for the central component of CR7 is not expected from either a Pop. III or DCBH model, which have SEDs that rapidly decline to longer wavelengths (S15;~\citealp{Agarwal2016}).
In addition, these models cannot reproduce the observed blue \IRACcolour colour which instead suggests the presence of significant rest-frame optical emission lines in the SED of CR7, including the {\sc [OIII]}$\,\lambda 4959, 5007$ line.

\section{Discussion}\label{sect:discussion}

\subsection{The Pop. III/DCBH candidate (A)}

\begin{figure}
\includegraphics[width = 0.45\textwidth]{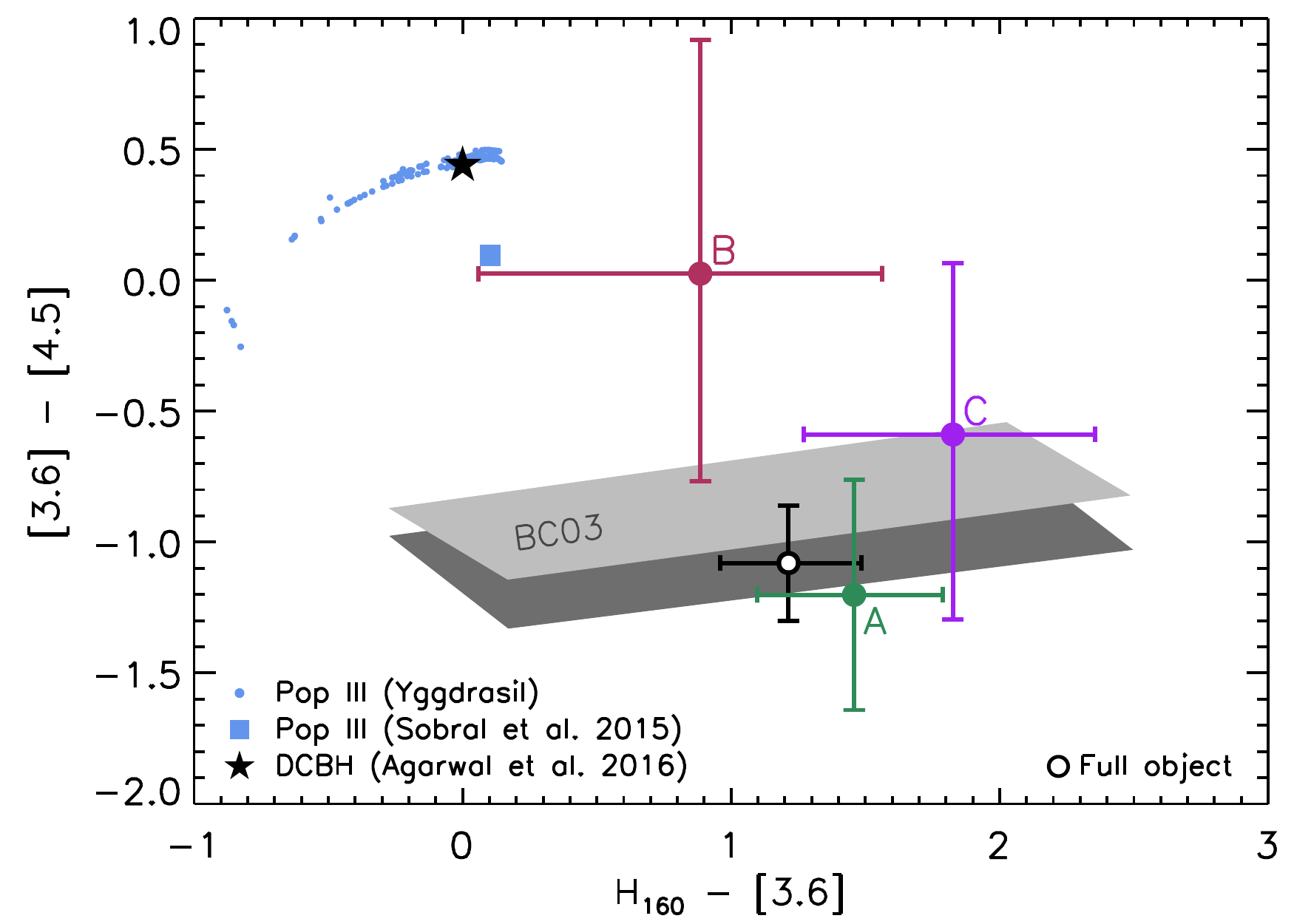}
\caption{The observed $H_{160} - [3.6]$ and \IRACcolour colours of the three individual components of CR7 (filled circles) and the full object (open circle), compared to predictions from both standard and exotic models.
We show the predicted colours of Pop. III models derived from {\sc Yggdrasil} (small blue circles) and from S15 (blue square).
The DCBH prediction from~\citet{Agarwal2016} is shown as the black star.
The grey shaded regions show the range of predicted colours from~\citet{Bruzual2003} models with ages of $10\, $--$250\,{\rm Myr}$ (effectively from left to right) where rest-frame optical emission lines have been added in the range $EW_{0}$\hboiii$ = 1000$--$2000$\AA.
The light grey region was calculated assuming a standard ratio of {\sc [OIII]}/{\sc H}$\beta$, whereas the dark grey region shows the results with {\sc [OIII]}/{\sc H}$\beta = 15$.}\label{fig:colours}
\end{figure}

As observed by S15, it is the brightest component (A) that is the source of the Lyman-$\alpha$ and \he emission and hence the Pop. III or DCBH candidate.
Using our deconfusion analysis, we find that component A has a strong detection in both~\emph{Spitzer} bands and shows a blue \IRACcolour colour.
This blue colour cannot be reproduced by pure Pop. III or DCBH models as shown in Fig.~\ref{fig:colours}, where we plot $H_{160} - [3.6]$ against \chone$-$\chtwo.
The figure shows the predicted colours of Pop. III models derived with the {\sc Yggdrasil} code~\citep{Zackrisson2011} and the best-fitting Pop. III model presented in S15.
The predicted colours from the {\sc Yggdrasil} code lie in a tight locus irrespective of the assumed star-formation history (burst or constant), initial mass function (IMF) of the Pop. III model or covering fraction ($f_{\rm cov} = 0.5$ and $1.0$).
We show the {\sc Yggdrasil} model results determined using a top-heavy IMF (`Pop. III.1' and `Pop. III.2' presented in~\citealp{Zackrisson2011}) and with ages less than $50\,{\rm Myr}$.
In addition, we show the expected colours of the best-fitting DCBH model from~\citet{Agarwal2016}.
The DCBH scenario similarly predicts low \chone~and \chtwo~fluxes and a flat \IRACcolour colour, in disagreement with the observations.

In contrast, the observed \IRACcolour colour agrees well with that typically measured for samples of Lyman-break galaxies at $z > 6$.
In these studies a blue colour of  \IRACcolour$\lesssim -1.0$ has been observed in both photometric samples~\citep{Smit2014, RobertsBorsani2015, Bowler2016}, and for galaxies with a spectroscopic confirmation~\citep{Finkelstein2013, Oesch2015, Stark2016}.
The origin of the blue rest-frame optical colour is believed to be contamination of these broad-bands by strong nebular-emission lines, 
with inferred rest-frame EW of $EW_{0}$\hboiii$ > 600$\AA~and potentially reaching in excess of $1800$\AA~\citep{Smit2014, Stark2016}.
For CR7 at $z = 6.6$, the most significant lines are the {\sc H}$\beta$ and {\sc [OIII]}$\,\lambda 4959, 5007$ emission lines in the~\chone~band, and the {\sc H}$\alpha$ line in the \chtwo~band (see Fig.~\ref{fig:sed}).
Given the canonical ratio of H$\alpha$ to H$\beta$ of $\sim 2.87$~\citep{Osterbrock2006}, we would expect to measure \IRACcolour $\gtrsim0.0$ if no emission from the {\sc [OIII]} line was present.
Instead we measure a strong blue colour, which is highly suggestive of the presence of the {\sc [OIII]} emission line.
In Fig.~\ref{fig:colours} we show the predicted colours from~\citet{Bruzual2003} SED models with the H$\alpha$, H$\beta$ and {\sc [OIII]} nebular emission lines added assuming a range of $EW_{0}$ consistent with that observed for high-redshift LBGs ($1000$\AA$\le EW_{0} \le 2000$\AA).
The SED models were created with both exponentially decreasing and constant star-formation histories in the age range from $10$ to $250\,{\rm Myr}$.
We assume no dust extinction here, as this would act to move the predicted colours away from the observed blue values.

The observed colours of CR7 can be reproduced (within the errors) with a young star-burst model with $EW_{0}$\hboiii$ \gtrsim 2000$\AA, assuming a standard [O{\sc III}]/H$\beta$ ratio for 1/5 $Z_{\sun}$ ($6.3$;~\citealp{Anders2003}).
With this ratio of [O{\sc III}]/H$\beta$, the observations for CR7 shown in Fig.~\ref{fig:colours} suggest an elevated rest-frame equivalent width of $EW_{0}$\hboiii$ \simeq 2500$\AA~would provide the best fit to the data.
Such a line-strength is higher than typically assumed at high redshift, but consistent with the observed colours of lensed galaxies found by~\citet{Smit2014}.
Alternatively, if the [O{\sc III}]/H$\beta$ ratio is in-fact larger at high redshift (as has been suggested in several studies;~\citealp{CurtisLake2013, DeBarros2016, Faisst2016}), then an even bluer \IRACcolour colour in agreement with that observed for component A can be obtained with $EW_{0}$ \hboiii $= 2000$\AA~(we assume [O{\sc III}]/H$\beta$ $= 15$;~\citealp{DeBarros2016}).
From this simple comparison, we find young ages ($< 100\,{\rm Myr}$) are required to reproduce the observed rest-frame UV to optical colour, $H_{160} - $\chtwo, for the brightest component of CR7 (A), which we find to be flat or slightly blue (at most the \chtwo~filter is boosted by $\Delta m_{\rm AB} \simeq 0.5$ due to H$\alpha$ with our assumed range in $EW_0$).
A young age would also be consistent with the observed rest-frame UV slope and the presence of the nebular lines of \he and \oiii, which typically require recent star-formation within the last $\sim 10\,{\rm Myr}$.
In conclusion, the observed rest-frame optical emission for CR7 measured with the~\emph{Spitzer} bands at \chone~and \chtwo~excludes a pristine Pop. III or DCBH model as presented in the current literature\footnote{Since the completion of this work, two additional studies have been presented,~\citealp{Agarwal2017} and~\citealp{Pacucci2017}, that claim to be able to reproduce the observed~\IRACcolour colour of CR7 with a DCBH model.}, and is instead typical of $z \simeq 7$ Lyman-break galaxies with strong emission by {\sc [OIII]}.

While the rest-frame optical emission of CR7 can be reproduced assuming nebular emission line strengths that are typical of $z \simeq 7$ LBGs, the presence of a strong and narrow \he emission line in the rest-frame UV spectrum of CR7 is unusual in observations of high-redshift LAEs and LBGs~\citep{Stark2014a, Stark2015, Stark2016, Zabl2015}.
To produce \he emission, a very hard ionizing spectrum is required~\citep{Stark2015}, and the large equivalent width of $EW_0 = 80 \pm 20$\AA~derived by S15 is challenging to reproduce with standard enriched stellar population model~\citep{Raiter2010, Schaerer2003}.
We find that the UltraVISTA DR3 broad-band photometric data shows a lower $J$-band excess, which implies a lower rest-frame equivalent width of \he of $EW_{0} = 40 \pm 30$\AA, or a line flux of $2.2^{+1.7}_{-1.6} \times 10^{-17}\,{\rm erg}/{\rm s}/{\rm cm}^2$, assuming the excess is purely a result of a single line.
While these values are consistent with the S15 measurements ($F_{\rm He} = 4.1 \pm 0.7 \times 10^{-17}\,{\rm erg}/{\rm s}/{\rm cm}^2$), the lower mean value (and more realistic errors) bring the measured $EW_{0}$ into the regime that is occupied by more standard interpretations such as AGN or emission from a low-metallicity star-burst.
We discuss these interpretations for the CR7 system in the next section.
\emph{HST} grism spectroscopy of CR7 (PI Sobral) will provide a more reliable estimate of the line flux and hence the equivalent width.

\subsection{The companion objects (B and C)}

Another important component of several of the Pop. III and DCBH models for CR7 is the presence of the supposedly older and redder B and C components at close proximity to the Pop. III/DCBH candidate (e.g.~\citealp{Agarwal2016, Pallottini2015}).
It is proposed that these companion galaxies provide the required Lyman-Werner radiation to prevent star formation at the site of CR7 and hence keep the gas pristine.
Both S15 and~\citet{Agarwal2016} claim that these components are red with best-fitting ages of $\gtrsim 300\,{\rm Myr}$.
Taking into account the considerable errors on the colours of components B and C, we find no evidence that they are particularly red in the rest-frame UV, however they do show a (relatively uncertain) red $H_{160} - $\chone~colour (see Fig.~\ref{fig:sed}).
\citet{Agarwal2016} claim that these components cannot be fitted with a single stellar population (SSP) model, however we find a large range of plausible star-formation histories (including SSPs) that can fit the observed photometry due to the sparse sampling in wavelength (5 filters) and the degeneracies present in SED fitting.
The lack of acceptable solutions in~\citet{Agarwal2016} appears to be a consequence of the assumption of no dust attenuation in the SED fitting analysis, however there is no evidence that these components are dust free.
We fit the $5$ band photometry ($I_{814}, \,YJ_{110}, \,H_{160}$, \chone, \chtwo) for components B and C using a range of SED models including dust attenuation with a maximum of $A_{\rm V} = 4.0$ assuming a~\citet{Calzetti2000} attenuation law, both constant and exponentially declining star-formation histories and a~\citet{Chabrier2003} IMF (full details are described in~\citealp{Bowler2014}).
For both components, the best-fitting model to the photometry is at $z \simeq 6.6$, however for component B a lower-redshift solution ($z = 1.4$) is formally acceptable.
Unsurprisingly given the large errors on the photometry for these faint components, we find that the ages and masses are considerably uncertain, with best-fitting ages of $200^{+600}_{-180}\,{\rm Myr}$ and $500^{+300}_{-350}\,{\rm Myr}$ for B and C respectively (computed by fixing the redshift to $z = 6.6$).
Note the large uncertainties on these derived ages, which span from several tens of ${\rm Myr}$ to the age of the Universe at $z = 6.6$, again a result of the large photometric errors and in particular the degeneracy between age and dust attenuation (e.g. see~\citealp{CurtisLake2013}).
The masses of components B and C are more constrained than the ages, with best-fitting values of ${\rm log}_{10}(M/{\rm M}_{\sun}) = 9.3^{+0.5}_{-0.6}$ and ${\rm log}_{10}(M/{\rm M}_{\sun}) = 9.7^{+0.2}_{-0.3}$ respectively.
These masses are considerably different to those determined in~\citet{Agarwal2016}, illustrating the uncertainties inherent in the SED fitting process, especially in the case of weak~\emph{Spitzer}/IRAC detections in confused imaging.

Finally, we perform SED fitting to the broad-band photometry available for the brightest component A of CR7, using only those filters that are uncontaminated by confirmed or probable, strong line emission.
The resulting fits prefer a low age $< 110\,{\rm Myr}$ ($1\sigma$), with a mass of ${\rm log}_{10}(M/{\rm M}_{\sun}) = 9.2^{+0.3}_{-0.1}$.
Note that the mass estimate here should be seen as an upper limit, as we have fitted to the observed \chtwo-band magnitude which is likely contaminated by {\sc H}$\alpha$ emission.
Calculating the SFR directly from the UV according to the~\citet{Madau1998} prescription, we find $SFR_{\rm UV} \simeq 20\,{\rm M}_{\sun}/{\rm yr}$.

\section{Alternative interpretations}\label{sect:alternative}

In this section we argue that more standard scenarios for the nature of CR7 are acceptable within the current observational constraints.
We consider here two alternative interpretations for CR7, the first being a low-mass, narrow-line AGN and the second being a young low-metallicity star-burst, modelled with the inclusion of binary stars.

\subsection{Type II AGN}

A natural explanation for the presence of \he emission is excitation by an AGN, whose power-law spectrum can extend into the rest-frame NUV and therefore provide the high-energy photons required to ionize Helium ($E = 54.4\, {\rm eV}$).
The spectra of AGN typically show fairly weak \he ($EW_{0}~\lesssim~10$\AA;~\citealp{Harris2016, VandenBerk2001}), however stronger emission lines of $EW_{0} \gtrsim 30$\AA~have been observed in samples of radio-selected and Type II objects at higher redshift (e.g.~\citealp{Matsuoka2009, Jarvis2005, Stern2002}).
S15 excludes an AGN based on the narrow line-width measured for He{\sc II}, the lack of metal lines measured in the near-infrared spectrum and the limits on the X-ray and radio luminosity.
While broad-line AGN cannot reproduce the width of the \he line in CR7 ($130 \pm 10\,{\rm km}\,{\rm s}^{-1}$), narrow-line obscured (Type II) AGN with low-mass black holes ($M_{\rm BH} \simeq 10^6\,{\rm M}_{\sun}$) are observed to have similarly narrow lines (e.g.~\citealp{Ludwig2012}), with \cite{Greene2005} showing that the narrow-line region traces the host galaxy velocity dispersion.
The lack of metal lines detected by S15 is also consistent with the range of line ratios exhibited by narrow-line AGN.
In particular, the lack of {\sc NV} emission has been observed in samples of Type II quasars from the Sloan Digital Sky Survey~\citep{Alexandroff2013} and in high-redshift radio galaxies~\citep{Matsuoka2009}.

The measured non-detection of the {\sc [CIII]}, {\sc CIII]}$\lambda\lambda1907,1909$ and {\sc OIII]}$\lambda \lambda 1661, 1666$ doublets with a ratio of He{\sc II}/{\sc CIII]}$ > 2.4$ and He{\sc II}/{\sc OIII]}$ > 2.4\, (1\sigma$; S15) is not sufficient to exclude an AGN either. 
Indeed, \cite{Feltre2016} show that such low ratios are more easily explained by the presence of an AGN rather than stellar processes.
There is a large spread in the observed line ratios for AGN that are consistent with the observations for CR7.
For example a similar {\sc CIII]}/\he ratio can be found in both radio-loud~\citep{Jarvis2001, Matsuoka2009} and {\em Spitzer-}selected AGN~\citep{MartinezSansigre2006}.
We note that from the available ground-based XSHOOTER spectrum, the prevalence of {\sc CIV} cannot be robustly constrained due to sky-line contamination at $1.18\mu {\rm m}$.
Type II AGN typically produce strong Lyman-$\alpha$ and rest-frame optical emission lines such as H$\alpha$, H$\beta$ and \oiii~in agreement with what we infer for CR7~\citep[e.g.][]{Zakamska2003}.
In-fact, the elevated \oiii/H$\beta$ ratios typically found for AGN~\citep[e.g.][]{Baldwin1981,Kewley2006} compared to galaxies, would result in a better agreement with the observed \IRACcolour colour for a similar assumed $EW_0$ (Fig.~\ref{fig:colours}).

\begin{figure}
\includegraphics[width = 0.22\textwidth]{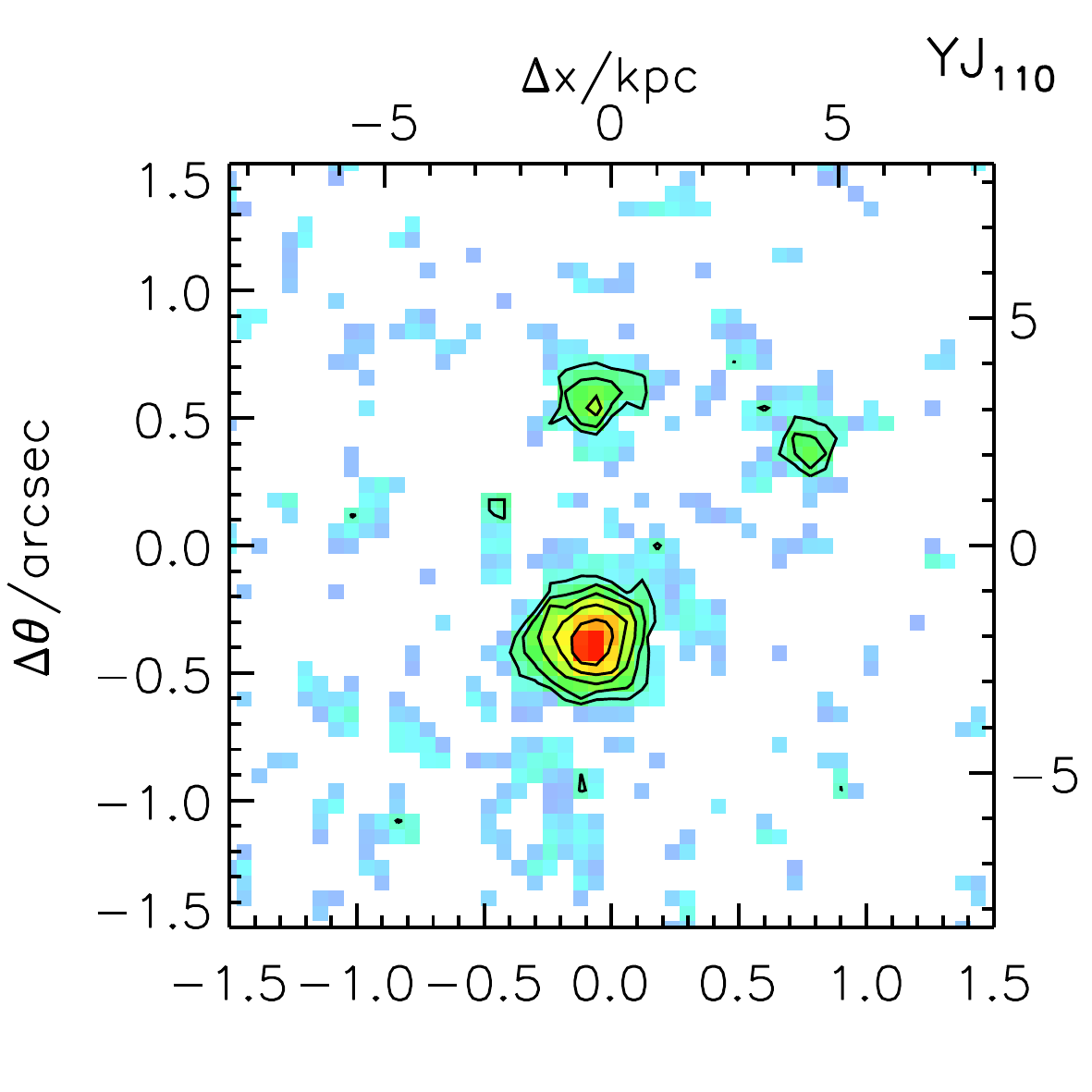}
\includegraphics[width = 0.22\textwidth, trim = 0 0.3cm 0 0]{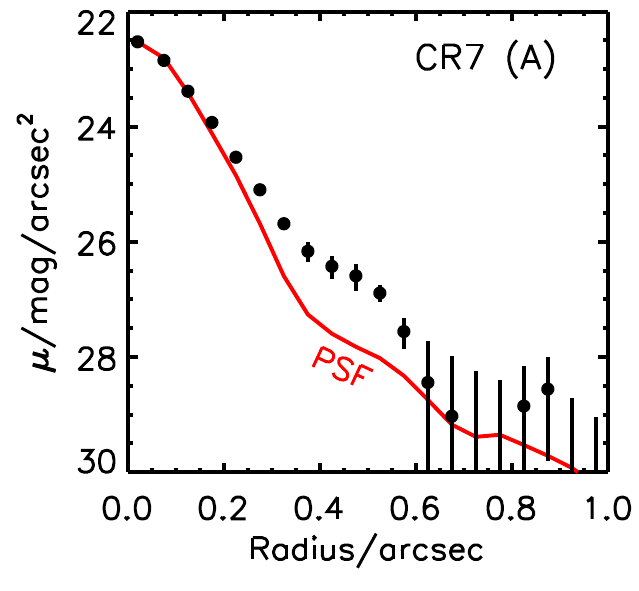}

\caption{The left-hand figure shows the $YJ_{110}$ data for CR7, scaled by surface brightness.
The contours trace $0.5$ mag intervals in the range $23.0$ to $25.5\,{\rm mag}^{-1}\,{\rm arcsec}^{-2}$.
The right-hand figure shows the surface-brightness profile of the brightest component of CR7 (component A), with the profile of the $YJ_{110}$ PSF shown as the solid red line.
}\label{fig:sbprofiles}
\end{figure}

The radio and X-ray imaging in COSMOS are insufficient to exclude a typical moderate-radio-luminosity AGN with the UV luminosity of CR7 (e.g. see a similar analysis for Himiko in~\citealp{Ouchi2009}).
The $5\sigma$ depth of the Very Large Array $1.4\,{\rm GHz}$ data in COSMOS is approximately $100\mu\,{\rm Jy}$ per beam at the position of CR7~\citep{Schinnerer2010}, and hence if CR7 had a similar ratio of UV to radio flux as shown in radio-loud quasars known at $z \simeq 6$ (e.g.~\citealp{Zeimann2011, McGreer2006}), it  would be only marginally detected in the available data.
Deeper Jansky-VLA imaging in the COSMOS field at $3\,{\rm GHz}$ (PI Smol{\v c}i{\' c}), that will extend a factor of 3 deeper than the current imaging, will provide greater constraints on the radio emission from CR7.
Finally, in the rest-frame UV, as probed at high resolution by the~\emph{HST}/WFC3 data, CR7 shows a compact morphology that is consistent with an AGN contribution.
Fig.~\ref{fig:sbprofiles} shows the surface-brightness profile of the brightest component of CR7 (A) derived from the $YJ_{110}$ data.
Extended flux is evident in the profile, however the central region is extremely compact and indistinguishable from a point source.
While the $YJ_{110}$ filter includes the Lyman-$\alpha$ emission line, we expect the surface brightness profile to be dominated by continuum emission.
This is because 1) the Lyman-$\alpha$ emission is considerably extended ($> 2 $ arcsec; S15) and 2) the emission line is at the very blue end of the photometric filter and hence the continuum flux dominates by around a factor of three.
A similarly compact surface-brightness profile is found in the shallower $H_{160}$ image.

\subsection{Low-metallicity star-burst}

In addition to an AGN, there are stellar sources such as Wolf-Rayet and hot massive stars that are capable of producing the hard UV photons required to ionize Helium (e.g.~\citealp{Eldridge2009,Kudritzki2002}).
While significant \he emission is uncommon in the local Universe, the presence of \he in the rest-frame UV spectra of $z = 2$--$4$ LBGs has been noted by several studies (e.g.~\citealp{Erb2010, Allam2007, Shapley2003}).
The typically broad \he lines observed, which show $EW_{0} \lesssim 7$\AA~\citep{Shapley2003, Cassata2013}, have been explained by the presence of Wolf-Rayet stars~\citep{Brinchmann2008} in these galaxies or alternatively through the consideration of binary star pathways~\citep{Eldridge2012}.
Using the BPASS models, both~\citet{Eldridge2012} and~\citet{Steidel2016} have found that by including massive binary stars in the SED modelling it is possible to reproduce the observed UV spectra of LBGs at $z = 2$--$4$ (including the diversity in the \he and C{\sc IV} emission-line strengths; also see~\citealp{Gutkin2016}).
In general, the BPASS models produce more ionizing radiation over a longer period than in binary-free models~\citep{Stanway2015}.
Binary stellar pathways lead to longer lived hot WR-like stars, which enhance the strength and longevity of the broad \he line~\citep{Eldridge2009} and furthermore lead to an elevated \oiii/H$\beta$ ratio as has been found in LBGs and Lyman-break analogues~\citep{Stanway2014, Steidel2016}.

\begin{figure}
\includegraphics[width = 0.48\textwidth, trim = 0 0.3cm 0cm 0.3cm]{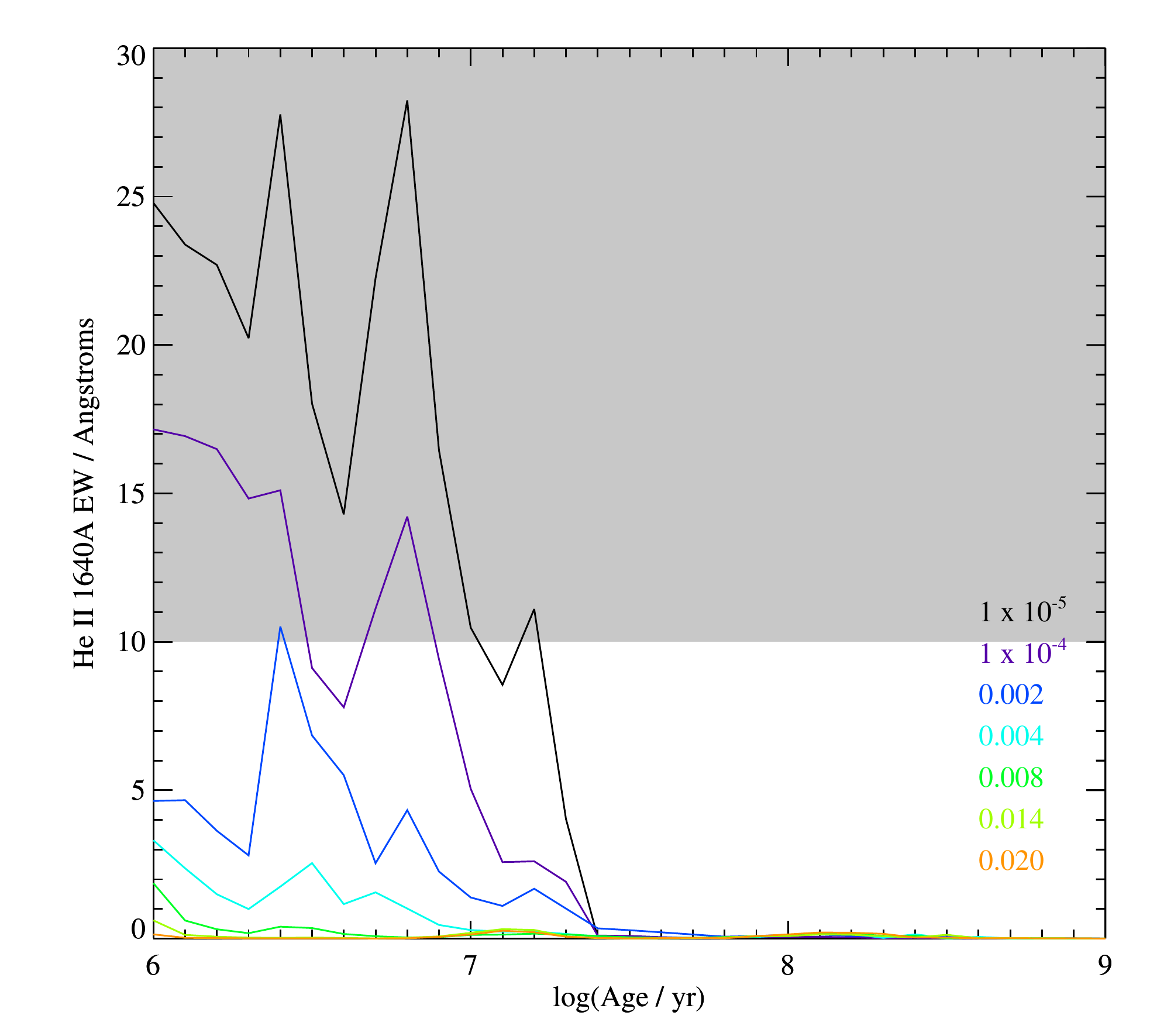}

\caption{The predicted \he equivalent width for the BPASS SED models described in the text, as a function of age.
The grey shaded region shows the $1\sigma$ lower limit on the $EW_{0}$ of this line, inferred from the UltraVISTA DR3 data studied here.
The coloured lines show a range of different metallicity mass fractions, which correspond to $\simeq\,{Z}_{\sun}$ to $1/2000\,{Z}_{\sun}$.
}\label{fig:bpass}
\end{figure}

In comparison to CR7 however, there are clear differences between the observed strong $EW_{0} = 80 \pm 20$\AA~(S15) and narrow ($\Delta v = 130 \pm 20\,{\rm km}/{\rm s}$) line and the \he emission in these lower-redshift LBGs.
Even with the lower inferred $EW_{0} = 40 \pm 30$\AA~that we find for CR7 in this study, the line strength is still significantly higher than that found in \he emitters observed at lower redshift, and is more compatible with an AGN origin~\citep{Cassata2013}.
In addition to the line strength, the width of the line is narrower than the stellar \he emission predicted from the BPASS models, which exceeds $1000\,{\rm km}/{\rm s}$ as a result of strong stellar winds from WR-type stars. 
In previous studies of the rest-frame UV emission lines produced by stellar population models that do not include binary stars, only with extremely low metallicities ($< 10^{-7}\,{\rm Z}_{\sun}$;~\citealp{Schaerer2003}, see also~\citealp{Raiter2010}) can the $EW_0$ value inferred for CR7 be produced.
Furthermore, in these models the \he emission is only present at very young ages, with the $EW_0$ dropping to essentially zero at ${\rm log}({\rm age}/{\rm yr}) > 6.3$ (fig. 7 in~\citealp{Schaerer2003}).
As noted in~\citet{Eldridge2012} and~\citet{Erb2010} however, the BPASS models presented previously do not include nebular line emission.
A nebular emission component produces a narrow line, often super-imposed on a broader component such as that observed for the $z = 2.3$ low-metallicity star-forming galaxy BX418 studied by~\citet{Erb2010}.
\citet{Cassata2013} also found a population of \he emitters with narrow line-widths, attributing these to either pockets of Pop. III star-formation or a peculiar stellar population.
Before we continue to compare the observed properties of CR7 to predictions of the nebular emission lines from BPASS, we note that an alternative source of the narrow \he emission observed in studies such as~\citet{Cassata2013} has been discussed recently by~\citet{Grafener2015}, who explore the emission from very massive stars (VMS).  
While we do not discuss such models further in this work, we note that with a moderately low metallicity ($0.01\,{\rm Z}_{\sun}$) the VMS models, which include WNh-type stars, can produce both strong \he emission with $EW_{0} \simeq 20$--$40$\AA ~(for single stars), line widths of a several hundred ${\rm km}/{\rm s}$ and weak metal lines.
These particular stellar pathways are not currently implemented in BPASS, and hence potentially their inclusion could further boost the predicted narrow \he component.

\begin{figure}
\includegraphics[width = 0.48\textwidth, trim = 0 0.3cm 0cm 0.3cm]{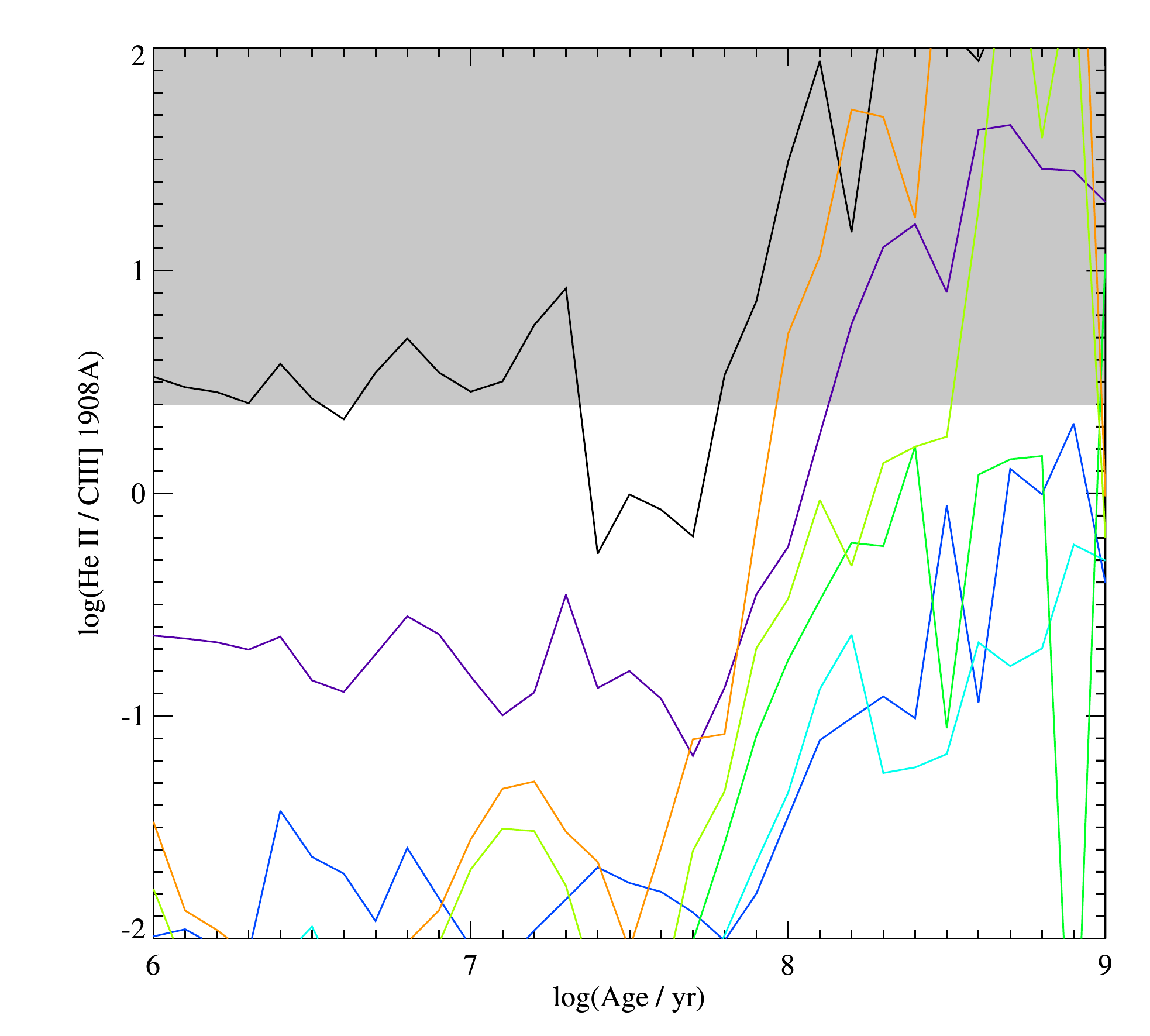}
\includegraphics[width = 0.48\textwidth, trim = 0 0.3cm 0cm 0.3cm]{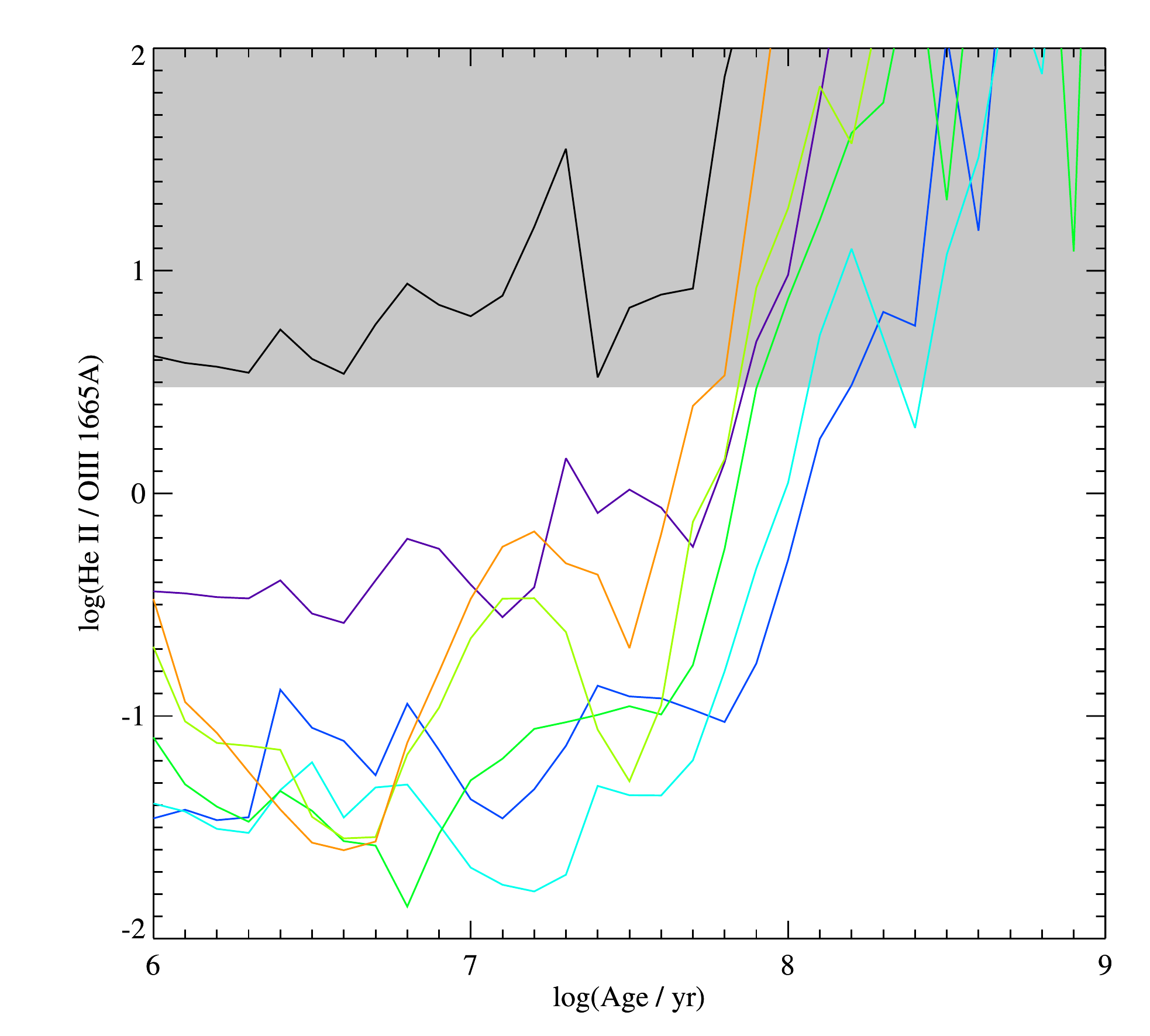}

\caption{In the upper and lower plot respectively, we show the \he to {\sc CIII]} and \he to {\sc OIII]} line ratio predicted by the BPASS models described in the text.
The grey shaded region in the lower plot shows the ratios allowed by the observations of CR7 presented in S15, represented by the $1\sigma$ upper limit in the {\sc CIII]} or {\sc OIII]} doublet flux.
Assuming the lower \he ${EW}_{0}$ found in this study, the limits drop to ${\rm log}_{10}($He{\sc II}/{\rm X}$) > 0.1 $($-0.2$) at $1\sigma\, (2\sigma)$ for both ${\rm X} =$ {\sc CIII]} and ${\rm X} =$ {\sc OIII]}.
As in Fig.~\ref{fig:bpass}, the coloured lines show a range of different metallicity mass fractions, which correspond to $\simeq\,{Z}_{\sun}$ to $1/2000\,{Z}_{\sun}$.
}\label{fig:bpass_limits}
\end{figure}

\subsubsection{Comparison to the BPASS models}

We investigated the potential strength of the nebular rest-frame UV and optical lines using the latest version of the BPASS models (v2.0; Eldridge et al. in prep.) coupled with radiative transfer effects modelled with {\sc CLOUDY}~\citep{Ferland2013}.
The models were generated using the same nebular gas geometry and physical conditions as presented in~\citet{Eldridge2009} and~\citet{Stanway2014}, with an electron density of $10^2 \,{\rm cm}^{-3}$.
A top-heavy initial mass function was used as described in~\citet{Stanway2015}, and the gas-phase metallicity was taken to be the same as the stellar metallicity.
The models were created assuming an instantaneous burst star-formation history.
In addition to the recently available BPASS v2.0 models, we include two models at lower metallicities ($1/200\,{Z}_{\sun}$ and $1/2000\,{Z}_{\sun}$) which are under development for release in BPASS v2.1.
In Fig.~\ref{fig:bpass} we show the expected nebular \he equivalent widths for these models with a range of metallicities (ranging from a mass fraction of $1 \times 10^{-5} $ to $ 0.02$, or equivalently $1/2000\,{Z}_{\sun}$ to ${Z}_{\sun}$ assuming a solar metallicity fraction of $0.02$).
The BPASS models shown at low-metallicity, predict significantly stronger nebular emission line strengths for \he than the broad stellar component (typically $1$--$3$\AA;~\citealp{Eldridge2012}).
In comparison to CR7, the models with young ages up to $\simeq 20\,{\rm Myr}$ and low-metallicities of $1/200$ or $1/2000\,{Z}_{\sun}$ are able to produce line emission consistent with the observed strength, showing $EW_{0} \simeq 10$--$30\,$\AA.
The $1/2000\,{Z}_{\sun}$ metallicity model is also consistent with the upper limits on the {\sc CIII]} and {\sc OIII]} emission lines determined observationally in S15 (the lower panels of Fig.~\ref{fig:bpass_limits}), although the higher-metallicity models struggle to reproduce this ratio at young ages.
We note that if the $EW_{0}$ of \he is lower than that determined by S15 as we find (see Section~\ref{sect:uv}), then this would imply that the upper limits on the He{\sc II}/C{\sc III}] and He{\sc II}/O{\sc III}] ratios are also a factor of two less stringent with ${\rm log}_{10}($He{\sc II}/{\sc CIII]}$) > 0.1 $($-0.2$) at $1\sigma\, (2\sigma)$ significance.
We therefore find that the BPASS models can reproduce the observed rest-frame UV lines of CR7, however we require that the system have a low-metallicity around $1/2000\,{\rm Z}_{\sun}$ given the range of input parameters that we explore in this study.

\begin{figure}
\includegraphics[width = 0.45\textwidth]{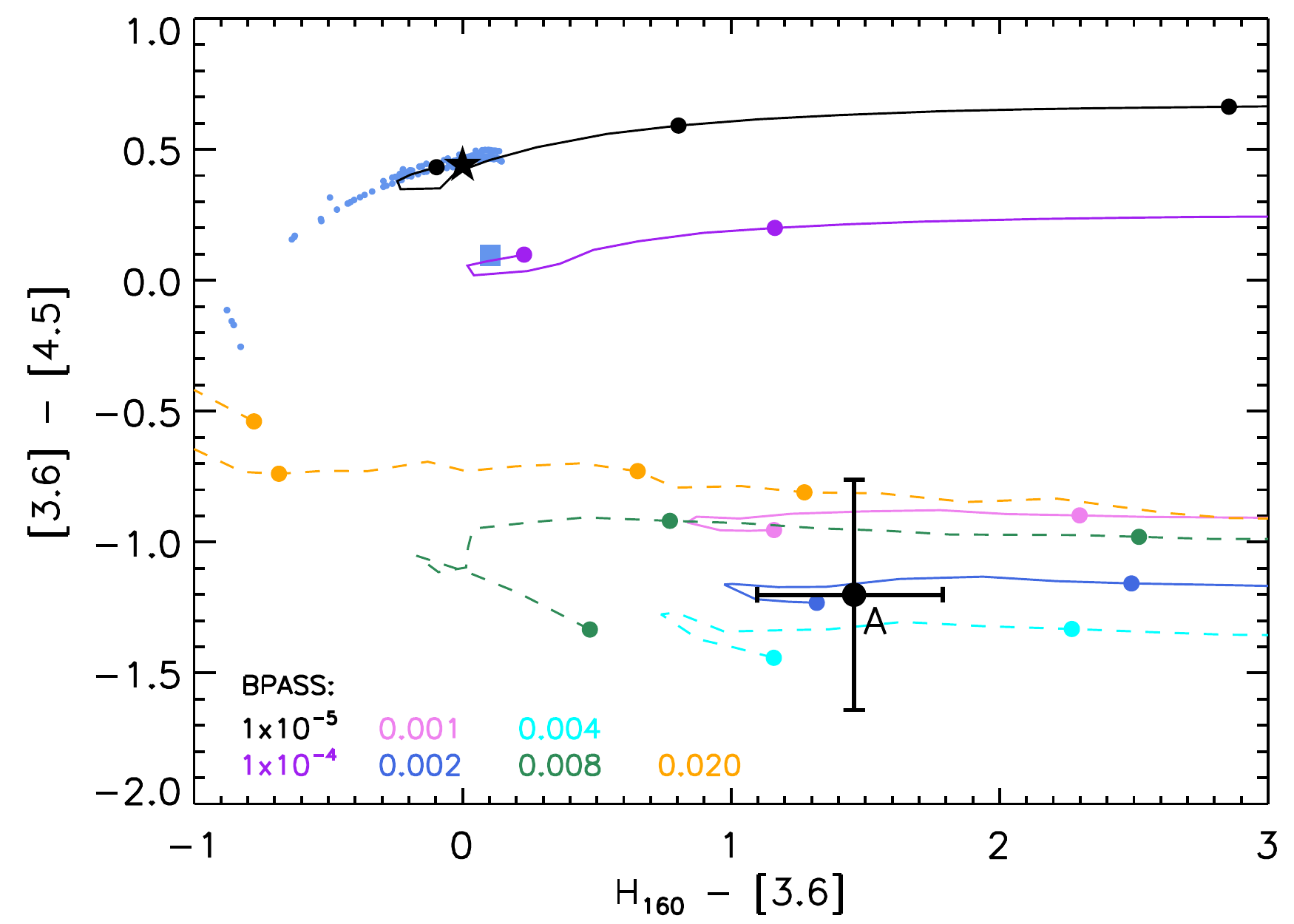}
\caption{The predicted $H_{160} - [3.6]$ and \IRACcolour colours of the BPASS models (lines), compared to the observed colours of the A component of CR7 (filled black circle).
The line colours specify the metallicity mass fraction of the model.
For each line the age increases from left to right, with circles denoting $1\,{\rm Myr}$, $10\,{\rm Myr}$, $50\,{\rm Myr}$ and $100\,{\rm Myr}$.
Dashed lines show metallicity mass fractions from $0.02$ to $0.004$ ($\simeq\,{Z}_{\sun}$ to $1/5\,{Z}_{\sun}$), which show a trend to a more negative \IRACcolour colour as the metallicity decreases.
As the metallicity drops further however, the trend is reversed (solid lines; $0.002$ to $1 \times 10^{-5}$ corresponding to $1/10\,{Z}_{\sun}$ to $1/2000\,{Z}_{\sun}$) and the predicted \IRACcolour colour eventually becomes positive.
The predicted colours of Pop. III and DCBH models are shown as the points in the upper right, as described in the caption of Fig.~\ref{fig:colours}.}\label{fig:colours_bpass}
\end{figure}

In addition to the rest-frame UV line strengths and ratios observed in CR7, any model of the system must be able to reproduce the strong emission in the rest-frame optical (from the Balmer and {\sc [OIII]} lines) that we infer from the~\emph{Spitzer}/IRAC photometry.
In Fig.~\ref{fig:colours_bpass} we show the $H_{160} - $\chone~and \IRACcolour colours predicted by the BPASS models we consider in this study, in comparison to the observed colour of the A component of CR7 (the Pop. III/DCBH candidate).
Above {10\,{\rm Myr}} the models show a constant \IRACcolour with age, as the H$\alpha$, H$\beta$ and {\sc [OIII]} emission lines decay with an approximately constant ratio after the initial burst.
The large variation in the $H_{160}$-\chone~colour with time is predominantly the result of the decay of the $H_{160}$ flux, which probes the far-UV wavelengths ($\lambda_{0} \sim 2000$\AA) at $z \simeq 7$.
At ages of a few Myr, the predicted colours vary on short timescales as the galaxy SED is dominated by a small number of the most massive stars, as they progress through the early stages of rotational mixing and quasi-homogeneous evolution~\citep{Eldridge2012}.

When comparing the predicted and observed rest-frame optical colours for CR7, we find that only models with metallicities of  $1/10 $--$1/20\,{\rm Z}_{\sun}$, and ages of $< 10\,{\rm Myr}$ can reproduce the observations for component A.
The required age of the models matches well with that required to reproduce the \he $EW_0$ and the rest-frame UV line ratios, however the metallicity needed to reproduce the \IRACcolour colour is higher by around a factor of $100$ than that required to produce sufficient \he flux (shown in Fig.~\ref{fig:bpass}).
This is a result of the lower metallicity models showing dramatically lower {\sc [OIII]} fluxes (e.g. as seen in the predicted {\sc [OIII]}/H$\beta$ ratios in~\citealp{Gutkin2016}), which leads to much redder \IRACcolour colours in comparison to CR7.
We therefore find that the BPASS models that we consider in this study, have difficulty reproducing simultaneously the observed strengths of the rest-frame UV and the rest-frame optical line strengths for CR7.

In recent work studying the rest-frame UV and optical spectra of $z \simeq 2.4$ LBGs however,~\citet{Steidel2016} found that to reproduce the full stellar and nebular spectra including the \he and {\sc [OIII]} emission line strengths, required a super-solar O/Fe ratio with O/Fe $\simeq 5\,$(O/Fe)$_{\sun}$.
The BPASS model predictions we show in Figures~\ref{fig:bpass},~\ref{fig:bpass_limits} and~\ref{fig:colours_bpass} were calculated based on Solar abundance ratios, and hence may not be an adequate representation of the conditions at high redshift.
To test the possible effect of such an increase in the $\alpha$-element abundances on the predicted line strengths from the BPASS models, we re-ran {\sc CLOUDY} after the gas-phase abundances of Mg, Ne, O and Si were increased by factors of $\sim 2$--$10$ times the solar values.
For an increase in the O/Fe ratio of $\simeq 10$ the predicted {\sc [OIII]} emission (and \IRACcolour colour) matches that of the BPASS model with a metallicity that is a factor of 10 higher.
Hence, the predicted \IRACcolour colours for the model with a stellar metallicity of $1/200\,{\rm Z}_{\sun}$ approximately follows that of the $Z_{\star} = 1/20\,{\rm Z}_{\sun}$ model, while the strength of the \he and Lyman-$\alpha$ emission are unaffected.
As is evident from Fig.~\ref{fig:colours_bpass}, the predicted rest-frame optical line ratios and subsequent \IRACcolour colour are very sensitive to metallicity between $1/20 $--$1/200\,{\rm Z}_{\sun}$, and hence an increase in the O/Fe abundance brings the BPASS models significantly closer to reproducing the \IRACcolour observed for CR7 (within $2\sigma$ for a factor of five increase in the O abundance).

In summary, with the particular parameters presented in this comparison, we cannot reproduce both the rest-frame UV and rest-frame optical emission line strengths with a BPASS (and {\sc CLOUDY}) model of a single common metallicity and assuming Solar element abundance ratios.
If however, at high-redshift there is significant $\alpha$-enhancement as expected for young systems, we find that a BPASS model with $Z_{\star} = 1/200\,{\rm Z}_{\sun}$ can reproduce the \he line strength ($EW_{0} > 10\,$\AA) and the inferred {\sc [OIII]} line strength (to within $2 \sigma$ assuming a five fold increase in the O/Fe ratio;~\citealp{Steidel2016}).
With such a model there is tension with the upper limits on the presence of other rest-frame UV lines as presented (for Solar abundance ratios) in Fig.~\ref{fig:bpass_limits}, however we caution that these lines can be sensitive to the particular physical conditions input into the photo-ionization code (e.g. see~\citealp{Feltre2016, Stanway2014}).
Given the uncertainties inherent in the modelling of nebular emission lines, we conclude that the BPASS models that include binary stars can provide a feasible galaxy model that broadly match the observational constraints available for CR7.
In particular, in comparison to other stellar evolution code predictions, only the BPASS models can currently produce the required nebular \he emission observed in CR7 without an AGN source.
With future independent measurements of the \he line strength, stronger upper limits on the presence of C{\sc III}] and other rest-frame UV lines, and the detection of rest-frame optical lines with the~\emph{James Webb Space Telescope (JWST)}, it will clearly be possible to more stringently constrain the stellar populations present in CR7.

\section{Conclusions}\label{sect:conc}

We provide improved constraints on the broad-band photometry for CR7 using deeper near-infrared imaging from the DR3 of the UltraVISTA survey, and deeper~\emph{Spitzer}/IRAC photometry from the SPLASH dataset.
The data show that the Pop. III/DCBH candidate in the CR7 system shows a strong, blue, rest-frame optical colour as measured by the~\emph{Spitzer}/IRAC \chone~and \chtwo~bands.
The magnitude and colour of these detections cannot be reproduced by the current Pop. III and DCBH models, and instead imply that the \chone~band is contaminated by the \oiii $\lambda \lambda 4959, 5007$ doublet with an inferred rest-frame equivalent width of $EW_{0}$\hboiii$ \gtrsim 2000$\AA.
Furthermore, the improved UltraVISTA DR3 near-infrared data show a lower $J$-band excess than previous studies, suggesting that the spectroscopically detected \he emission line has a lower inferred rest-frame equivalent width of $EW_{0} = 40 \pm 30$\AA.
The observational constraints on the \he and {\sc [OIII]} emission line strengths are consistent with the properties of a narrow-line low-mass AGN or, alternately, a young low-metallicity $\sim 1/200\,{\rm Z}_{\sun}$ star-burst when modelled including binary stars and an enhanced O/Fe abundance ratio.
However, we find that this star-burst model (from the BPASS code) cannot reproduce the current upper limits on the lack of metals in the near-infrared (rest-frame UV) spectrum.
In contrast, such ratios of {\sc CIII]}/\he and {\sc OIII]}/\he are to be expected for an AGN source~\citep{Feltre2016}.
Future observations of CR7 (and other high-redshift galaxies with likely strong rest-frame optical emission) with~\emph{JWST} will be able to directly detect the inferred \oiii~emission line at $\lambda_{\rm obs} \simeq 3.8\,\mu {\rm m}$, and through the measured line ratios and widths, will be able to distinguish between an AGN or a low-metallicity star-forming galaxy.

\section{Acknowledgements}
This work was supported by the Oxford Centre for Astrophysical Surveys which is funded through generous support from the Hintze Family Charitable Foundation.
JSD acknowledges the support of the European Research Council via the award of an Advanced Grant (PI J. Dunlop), and the contribution of the EC FP7 SPACE project ASTRODEEP (Ref.No: 312725). 
RJM and DJM acknowledge the support of the European Research Council via the award of a Consolidator Grant (PI R. McLure).
ERS acknowledges support from UK Science and Technology Facilities Council (STFC) consolidated grant ST/L000733/1.
MJJ acknowledges support from the UK STFC [ST/N000919/1].
This publication arises from research partly funded by the John Fell Oxford University Press (OUP) Research Fund
ERS and JJE wish to acknowledge the contribution of the high-performance computing facilities and the staff at
the Centre for eResearch at the University of Auckland.
New Zealand's national facilities are provided by the New Zealand eScience Infrastructure (NeSI) and funded jointly
by NeSI's collaborator institutions and through the Ministry of Business, Innovation and Employment Infrastructure programme (\url{http://www.nesi.org.nz}).
This work is based in part on observations made with the NASA/ESA~\emph{Hubble Space
Telescope}, which is operated by the Association of Universities for Research in Astronomy, Inc, under NASA contract
NAS5-26555. 
This work is based on data products from observations made with ESO Telescopes at the La Silla Paranal Observatories under ESO programme ID 179.A-2005 and on data products produced by TERAPIX and the Cambridge Astronomy survey Unit on behalf of the UltraVISTA consortium.
This work is based in part on observations made with the~\emph{Spitzer Space Telescope}, which is operated by the Jet Propulsion Laboratory, California Institute of Technology under a NASA contract.

\bibliographystyle{mnras} 
\bibliography{library_abbrv.bib}

\appendix

\section{IRAC deconfusion without component A}

\begin{figure*}
\includegraphics[width = \textwidth]{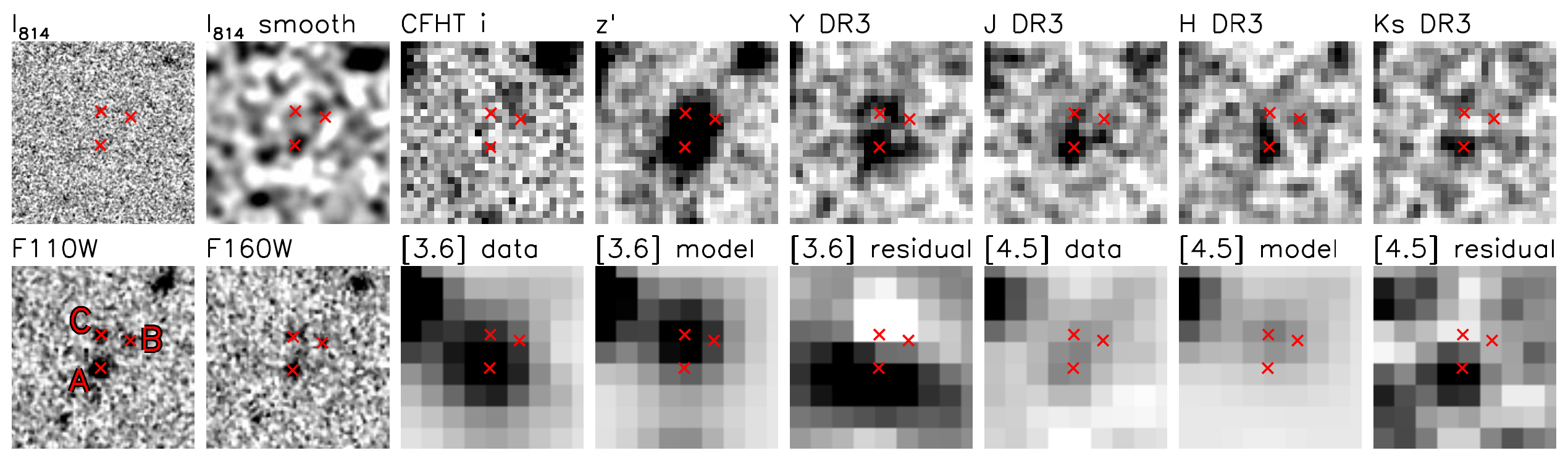}
\caption{Postage-stamp cut-out images for CR7.
Here we show the results of the deconfusion analysis of the~\emph{Spitzer}/IRAC images where no contribution from component A is allowed.
The size and scaling of the stamps is as described in the caption to Fig.~\ref{fig:photo}.
For the IRAC stamps of the data and model, we saturate pixels to black if they exceed the $10\sigma$ limit ($\sigma$ here is per pixel), and to white if they are less than $-1\sigma$.
The residual stamps are scaled in the same way but in the range of [$-2\sigma$, $2\sigma$].
}\label{fig:photo_noA}
\end{figure*}

In Fig.~\ref{fig:photo_noA} we show the results of the deconfusion analysis of the~\emph{Spitzer}/IRAC data if the A component is excluded from the modelling.
This situation is advocated by S15, who predict low fluxes in the IRAC bands from component A (the Pop. III candidate).
Without component A, our deconfusion model works by adjusting the contributions to the \chone~and \chtwo~bands at the positions of component B and C, in an attempt to match the data.
Comparing Fig.~\ref{fig:photo} and Fig.~\ref{fig:photo_noA}, it can be seen that excluding component A results in significantly larger residuals.
This is particularly clear in the \chone-band, where the best-fitting model without component A is unable to reproduce the data well, resulting in a residual of $\simeq 50$ percent of the peak flux.

\bsp	
\label{lastpage}
\end{document}